\documentclass[%
 reprint,
 showpacs,
 amsmath,amssymb,
 aps,
 pra,
]{revtex4-1}

\usepackage[ascii]{inputenc}
\usepackage[T1]{fontenc}
\usepackage[english]{babel}
\usepackage{graphicx}
\usepackage{microtype}
\usepackage[pdftex,colorlinks]{hyperref}
\hypersetup{%
  linkcolor=blue,%
  citecolor=blue,%
  urlcolor=blue,%
  pdftitle={PT-symmetric currents of a Bose-Einstein condensate in a triple
  well},%
  pdfauthor={ Daniel Haag, Dennis Dast,
    Holger Cartarius, G\"unter Wunner}%
}

\DeclareMathOperator{\trace}{tr}

\newcommand{\PT}{\mathcal{PT}}

\newcommand{\mrm}{\mathrm}
\newcommand{\rmi}{\mathrm{i}}

\newcommand{\bra}[1]{\langle{#1}|}
\newcommand{\ket}[1]{|{#1}\rangle}

\begin{document}

\title{$\PT$-symmetric currents of a Bose-Einstein condensate in a triple well}

\author{Daniel Haag}
\email[]{daniel.haag@itp1.uni-stuttgart.de}

\author{Dennis Dast}

\author{Holger Cartarius}

\author{G\"unter Wunner}

\affiliation{Institut f\"ur Theoretische Physik 1,
  Universit\"at Stuttgart, 70550 Stuttgart, Germany}

\date{\today}

\begin{abstract}
  We study the case of $\PT$-symmetric perturbations of Hermitian Hamiltonians
  with degenerate eigenvalues using the example of a triple-well system.
  The degeneracy complicates the question, whether or not a stationary current
  through such a system can be established, i.e.\ whether or not the
  $\PT$-symmetric states are stable.
  It is shown that this is only the case for perturbations that do not couple
  to any of the degenerate states.
  The physical explanation for the inhibition of stable currents is discussed.
  However, introducing an on-site interaction restores the capability to
  support stable currents.
\end{abstract}


\maketitle

\section{Introduction}
\label{sec:Introduction}
Bose-Einstein condensates (BECs) of dilute ultracold atom gases are one of the
most prominent examples of many-particle systems that can be described by a
mean-field theory.
For temperatures that are considerably smaller than the critical temperature of
the condensate the Gross-Pitaevskii equation~\cite{Gross61a,Pitaevskii61a} is
known to provide very accurate results, although the description has
limitations in the vicinity of dynamic instabilities~\cite{Anglin01a,
Vardi01a}.

There is a broad field of applications of such BECs in fundamental
research, covering anything from the simulation of condensed matter in optical
lattices over the study of rotating ultracold gases to the transition from
fermionic superfluids to BECs~\cite{Bloch08a}.
Another promising application is the atom laser~\cite{Holland96a}.
To realize a stable source of coherent atoms for the study of transport
phenomena or the realization of an atom laser, particles have to be coupled in
and out of the condensate coherently.
Both has already been realized experimentally~\cite{Gericke08a, Robins08a}.

To make use of the Gross-Pitaevskii equation even in the case of particle
exchange with the environment, one can turn to non-Hermitian
systems~\cite{Moiseyev11a}.
Here particle in- and outflux are described by positive and negative imaginary
potentials, respectively~\cite{Kagan98a}.
Exact many-particle calculations~\cite{Rapedius13a} and master equation
approaches~\cite{Rapedius13a, Dast14a,Trimborn08a, Witthaut11a} support the use
of the non-Hermitian mean-field ansatz.
Other applications of non-Hermitian Hamiltonians in dissipative optical
lattices~\cite{Abdullaev10a, Bludov10a} and in the study of condensate
decay~\cite{Rapedius09a, Rapedius10a}, as a result of a complex scaling
approach~\cite{Schlagheck06a}, have also been very successful.

However, it is not quite clear whether these Hamiltonians provide really
stationary states supporting a stable current through the system since the
energy eigenvalues can take complex values.
A condition that is known to provide real eigenvalues in many cases is $\PT$
symmetry~\cite{Bender98a, Bender99a, Bender07a}.
Such systems have either real eigenvalues, belonging to $\PT$-symmetric
eigenstates, or complex conjugate eigenvalues, belonging to $\PT$-broken
eigenstates.
The concept of $\PT$ symmetry is a special case of the a theory of
pseudo-Hermitian operators~\cite{Mostafazadeh02a, Mostafazadeh02b,
Mostafazadeh02c}, however, both properties become identical in the case of a
Schr\"odinger equation in position space with complex potentials.

The capability of these systems to provide real eigenvalue spectra was analyzed
in many theoretical studies~\cite{Klaiman08a, Schindler11a, Bittner12a,
Cartarius12a, Cartarius12b, Mayteevarunyoo13a, Graefe08b, Graefe12a} and was
realized experimentally in optical wave-guide systems~\cite{Klaiman08a,
Ruter10a, Guo09a, Peng14b}, which promises technical applications.
Even though proposals exist~\cite{Kreibich13a,Single14a}, the concept has not
yet been realized experimentally in ultra-cold atomic gases.

These concepts have in common that a Hermitian system is perturbed by a 
non-Hermitian $\PT$-symmetric Hamiltonian. 
For weak perturbations such systems can be discussed using perturbation
theory.
This technique has already been used with great
success~\cite{Fernandez98a,Fernandez14a} to decide whether a linear system
stays $\PT$ symmetric under a non-Hermitian perturbation.
Since complex eigenvalues lead to an exponential growth or decay of the
eigenstate, $\PT$-symmetric states in a linear system are stable if and only if
the eigenvalue spectrum is entirely real, i.e.\ the symmetry is unbroken.

We want to tackle the question under which conditions a system, in particular
a three-well setup, is capable of supporting a stable stationary current, i.e.\
stable stationary states for finite $\PT$-symmetric imaginary potentials. 
For simple, i.e.\ non-degenerate, discrete eigenvalues which become equally
spaced for high quantum numbers, it has already been shown that real
eigenvalues always exist and can be found if the non-Hermitian part of the
Hamiltonian is small enough~\cite{Mityagin13a,Adduci12a,Haag14a}.
However, this situation can change completely if degenerate eigenvalues are
considered.
The focus of the present work lies on the question how such degeneracies
influence the existence of stable $\PT$-symmetric states.
Additionally, since we are interested in BECs, the impact of the nonlinear
contact interaction in the Gross-Pitaevskii equation on the linear results will
be studied.
In this case the stability has to be investigated separately using the
Bogoliubov-de Gennes equations.

In Sec.~\ref{sec:PerturbationTheory}, we recapitulate the perturbation theory
approach for both anti-Hermitian and pseudo-Hermitian perturbations and derive
a stringent condition under which the $\PT$ symmetry is preserved at least for
smallest perturbations.
Then we turn to a three-well system whose properties are discussed in the
linear and nonlinear case in Sec.~\ref{sec:ThreeMode}, testing the prediction
from the perturbation theory approach and the influence of the inter-particle
interaction.
Finally conclusions are drawn in Sec.~\ref{sec:Conclusion}.

\section{Perturbation Theory}
\label{sec:PerturbationTheory}
We discuss a linear system described by the Hamiltonian $\hat{H} =
\hat{H}_\mrm{0}+\gamma\hat{H}_\mrm{P}$, where the unperturbed Hamiltonian
$\hat{H}_\mrm{0}$ is Hermitian and the perturbation $\hat{H}_\mrm{P}$ is
anti-Hermitian.
If they fulfill the relations
\begin{subequations}
\begin{align}
  \mathcal{P} \hat{H}_\mrm{0} = \hat{H}_\mrm{0}^\dagger \mathcal{P}
                              = \hat{H}_\mrm{0} \mathcal{P},\\
  \label{eq:pertHamiltonian}
  \mathcal{P} \hat{H}_\mrm{P} = \hat{H}_\mrm{P}^\dagger \mathcal{P}
                              = - \hat{H}_\mrm{P} \mathcal{P},
\end{align}
\end{subequations}
the Hamiltonian is pseudo-Hermitian with respect to the parity operator,
$\mathcal{P}\hat{H} = \hat{H}^\dagger \mathcal{P}$.
It is obvious that in case of a diagonal matrix representation, the
system is $\PT$ symmetric, where $\mathcal{T}$ acts as a complex conjugation.
This is in particular true in position space where the perturbation is given by
an antisymmetric imaginary potential $\rmi\hat{V}\left(\hat{x}\right)$.

If all eigenvalues of $\hat{H}_\mrm{0}$ are non-degenerate and discrete a
perturbation can always be chosen small enough such that the eigenvalues,
provided they behave as continuous functions of the perturbation strength
$\gamma$, do not coalesce and form exceptional points.
Since complex eigenvalues of pseudo-Hermitian Hamiltonians always appear in
complex conjugate pairs~\cite{Mostafazadeh02a, Mostafazadeh02b,
Mostafazadeh02c} the formation of complex eigenvalues can only occur at
exceptional points.
In particular, for perturbations small enough all eigenvalues remain real.

This particular property can also be seen in a more formal way using
perturbation theory~\cite{Fernandez01a, Messiah65a}.
In addition, this approach can later be used to analyze degenerate systems.
Since the unperturbed Hamiltonian commutes with the parity operator
$\mathcal{P}$, all eigenvectors, $\{\ket{\psi_m}\}$, can be chosen either
symmetric or antisymmetric, hence $\mathcal{P} \ket{\psi_m} = \lambda_m
\ket{\psi_m}, \lambda_m = \pm 1$ holds.
The power series for the energy eigenvalue $\mu_n$ of the corresponding
eigenstate $\ket{\psi_n}$ reads
\begin{equation*}
  \mu_n = \mu_{n,0} + \gamma \mu_{n,1} + \gamma^2 \mu_{n,2} + \dots.
\end{equation*}
A useful explicit form of the energy corrections was given by Tosio Kato
in 1949~\cite{Kato49a},
\begin{equation}
  \mu_{n,s} = (-1)^{s-1} \trace \sum\limits_{\mathbf{k}_s}
  S^{k_1} \hat{H}_\mrm{P} S^{k_2} \cdots  \hat{H}_\mrm{P} S^{k_{s+1}},
  \label{eq:pertExplicitEnergy}
\end{equation}
where the sum is chosen to iterate over all $\mathbf{k}_s \in \mathbb{N}^{s+1}$
with $k_1+\ldots+k_{s+1} = s-1$.
The {\it reduced resolvent} is defined as
\begin{equation*}
  S = \sum\limits_{m \neq n} \frac{\ket{m}\bra{m}}{\mu_m - \mu_n},
\end{equation*}
while the zeroth power is treated differently as
\begin{equation*}
  S^0 = \ket{n}\bra{n}.
\end{equation*}
We see that $[S,\mathcal{P}] = 0$ since
\begin{equation*}
  \mathcal{P} \ket{m}\bra{m} = \lambda_m \ket{m}\bra{m} =  \ket{m}\bra{m}\lambda_m
  = \ket{m}\bra{m} \mathcal{P}.
\end{equation*}

Pseudo-Hermiticity of $\hat{H}_\mrm{P}$ ensures that all corrections
$\mu_{n,s}$ are real.
This can be seen by using $\trace(\mathcal{P} \hat{O} \mathcal{P}) = \trace
(\hat{O})$ and $S^\dagger = S$ and calculating the complex conjugate of the
energy correction
\begin{align}
  \mu_{n,s}^* &= (-1)^{s-1} \trace \sum\limits_{\mathbf{k}_s}
  \mathcal{P} S^{k_{s+1}} \hat{H}_\mrm{P}^\dagger \cdots
  \hat{H}_\mrm{P}^\dagger S^{k_{1}}\mathcal{P} \notag \\
  &= (-1)^{s-1} \trace \sum\limits_{\mathbf{k}_s}
  S^{k_{s+1}} \hat{H}_\mrm{P} \mathcal{P}  \cdots \mathcal{P}
  \hat{H}_\mrm{P} S^{k_{1}} \notag \\
  &= (-1)^{s-1} \trace \sum\limits_{\mathbf{k}_s}
  S^{k_{1}} \hat{H}_\mrm{P} \cdots 
  \hat{H}_\mrm{P} S^{k_{s+1}} \notag \\
              &=\mu_{n,s}.
  \label{eq:pertCorrectReal}
\end{align}
However, since anti-Hermiticity of $\hat{H}_\mrm{P}$ renders every odd energy
correction purely imaginary,
\begin{align}
  \mu_{n,s}^* &= (-1)^{s-1} \trace \sum\limits_{\mathbf{k}_s}
  S^{k_{s+1}} \hat{H}_\mrm{P}^\dagger \cdots
  \hat{H}_\mrm{P}^\dagger S^{k_{1}} \notag \\
  &= (-1)^{s-1} \trace \sum\limits_{\mathbf{k}_s}(-1)^s
  S^{k_1} \hat{H}_\mrm{P} \cdots  \hat{H}_\mrm{P} S^{k_{s+1}} \notag \\
  &= (-1)^{s} \mu_{n,s},
  \label{eq:pertCorrectImag}
\end{align}
they vanish and we conclude that the lowest order energy shift is at least
quadratic.

The situation changes drastically if we stop assuming non-degenerate
eigenvalues. 
Let us consider $M$ degenerate eigenvectors $\{\phi_i\}$ with the common
eigenvalue $\mu$.
The first order energy shift is given by the eigenvalues of the matrix
\begin{equation}
  S_{ij} =
  \bra{\phi_i}\hat{H}_\mrm{P}\ket{\phi_j}.%
  \label{eq:PerturbationMatrix}
\end{equation}%
The matrix $S$ inherits the anti-Hermiticity from the perturbation
$\hat{H}_\mrm{P}$.
Its eigenvalues are therefore either purely imaginary or zero.
This means that any perturbation that does not imply $S = 0$ immediately
produces imaginary eigenvalues.
For $\PT$-symmetric Hamiltonians describing in- and outcoupling of particles,
this means that no stable stationary current from particle source to particle
sink can exist.

For $S=0$, i.e.\ all matrix elements between degenerate eigenstates are zero,
higher-order perturbations must be taken into account.
In this case the considerations for simple eigenvalues from
Eq.~\eqref{eq:pertCorrectReal} hold, thus all eigenvalues remain real.

\section{Three-mode system}
\label{sec:ThreeMode}
To gain a more physical understanding of the effects leading to this symmetry
breaking that prohibits any stable stationary current through a system, we turn
to a simple example and solve the three-mode GPE with on-site interactions,
\begin{equation}
  \sum\limits_{j=1}^{3} H_{ij} \psi_j + U \left| \psi_i \right|^2
  \psi_i  = \mu \psi_i,
  \label{eq:ThreeModeGpe}
\end{equation}%
described by the Hamiltonian
\begin{equation}
  \mathbf{H} = 
  \underbrace{
    \begin{pmatrix}
    0 & -J & -1 \\
    -J & 0 & -J \\
    -1 & -J & 0
    \end{pmatrix}
  }_{\hat{H}_0}
  +\gamma 
  \underbrace{
    \begin{pmatrix}
    \rmi & 0 & 0 \\
    0 & 0 & 0 \\
    0 & 0 & -\rmi 
    \end{pmatrix}
  }_{\hat{H}_\mrm{P}}
  .
  \label{eq:ThreeModeHamiltonian}
\end{equation}
The parameter $U$ is the strength of the nonlinear on-site interaction.
The chemical potential $\mu$ takes the role of the energy and, for the linear
case, $U=0$, is given by the eigenvalues of the Hamiltonian.
Therefore the three terms are used as synonyms.

\begin{figure}%
  \centering%
  \includegraphics[width=0.5\columnwidth]{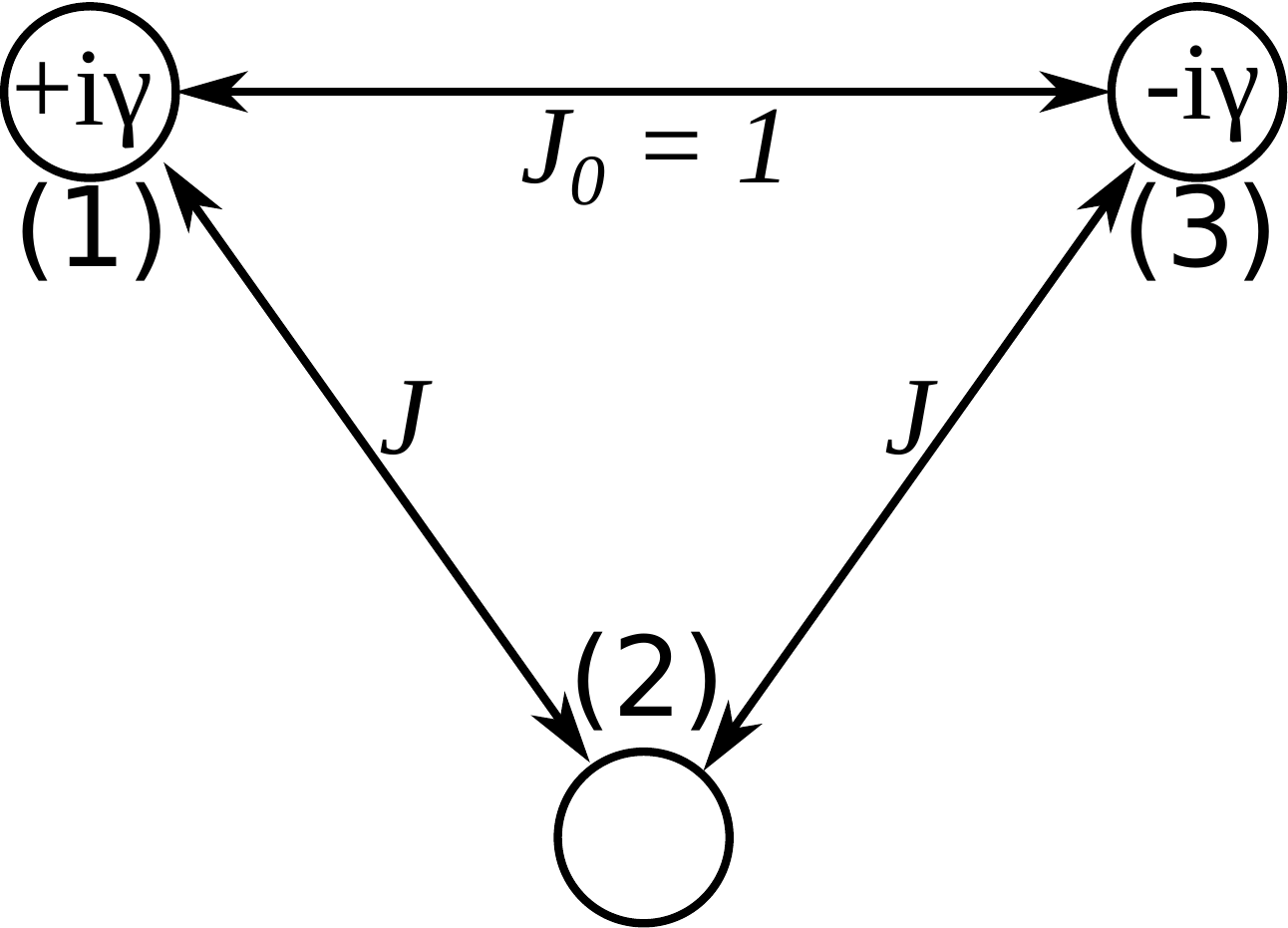}%
  \caption{%
    Scheme of the three-mode model. The imaginary on-site energies provide
    a particle influx to well 1 and an outflux from well 3. These two wells
    are coupled by a constant strength, set to unity, while the coupling $J$ to
    well 2 is varied.
  }%
  \label{fig:System3Mode}%
\end{figure}

The Hamiltonian describes a system of three coupled wells where in the first
well particles are injected into the system while particles are removed from
the third well at the same rate $\gamma$.
The tunneling rate between the gain and loss well is fixed to unity while the
second well is coupled to both by the same variable coupling strength $J$.
A scheme of the setup is shown in Fig.~\ref{fig:System3Mode}.

It is obvious that this Hamiltonian is $\PT$-symmetric and pseudo-Hermitian
with respect to
\begin{equation*}
  \mathcal{P} = \begin{pmatrix}0&0&1\\0&1&0\\1&0&0\end{pmatrix}.
\end{equation*}

The effects leading to the symmetry breaking introduced in the
previous section are discussed in the linear system, $U=0$. 
Without the perturbation, i.e.\ $\gamma=0$, the eigenvalues are given by the
simple expressions
\begin{align}
  \mu_1 &= -\sqrt{2J^2+1/4}-1/2, \\
  \mu_2 &= +\sqrt{2J^2+1/4}-1/2, \\
  \mu_3 &= 1.
  \label{eq:ThreeModeGamma0}
\end{align}
There are two cases that deserve special attention.
For $J=0$ the second well is not coupled to the remaining double-well system.
This case has already been discussed in detail~\cite{Graefe12a}.
The $\PT$ symmetry in the system breaks at $\gamma = 1$.
The second special case is $J=1$, in which the system becomes totally
symmetric.
In this situation the two eigenvalues $\mu_2 = \mu_3 = 1$ degenerate.
Applying perturbation theory sketched in the previous section to this case
leads to the matrix
\begin{equation}
  S = \begin{pmatrix}
    0 & 2\rmi \\
    2\rmi & 0 
    \end{pmatrix},
  \label{eq:ThreeModeJ1PertMatrix}
\end{equation}
immediately breaking the $\PT$ symmetry.

\begin{figure}%
  \centering%
  \includegraphics[width=\columnwidth]{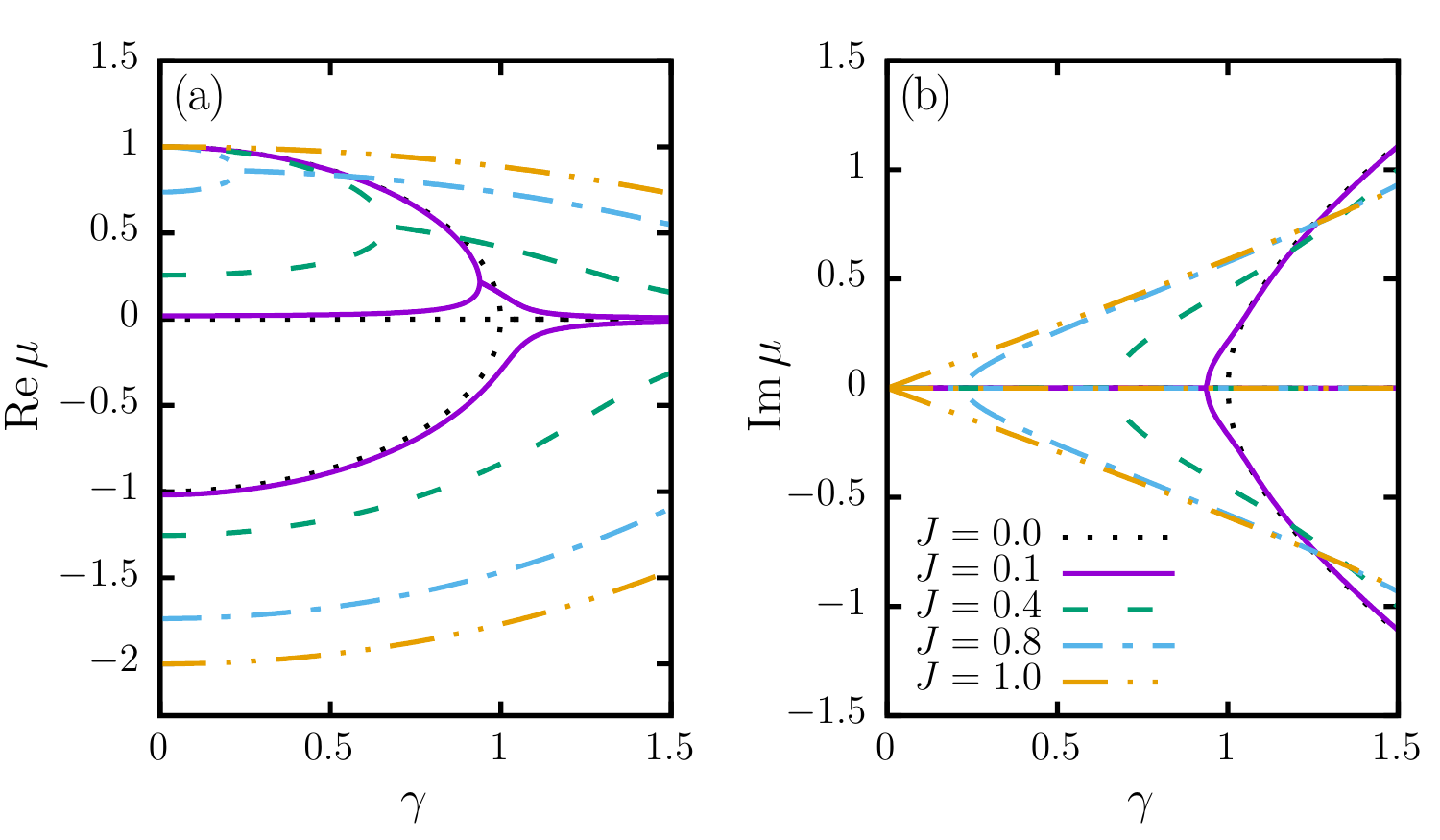}%
  \caption{%
    (Color online)
    (a) Real and (b) imaginary part of the eigenvalue $\mu$ for different
    values of the coupling strength $J$ as function of the in- and outcoupling
    parameter $\gamma$.
    For $J=0$ the ground and second excited state form an exceptional
    point of second order (EP2) at $\gamma=1$ where both $\PT$-symmetric states
    vanish and two $\PT$-broken states emerge.
    The first excited state, located only in well 2 is completely independent
    from the other eigenstates. 
    For $J>0$ the first excited state forms an EP2 with the second excited
    state, while the ground state is shifted to smaller energies and retains
    its $\PT$ symmetry for all values of $\gamma$.
    For increasing values of $J$, the energy difference between the two excited
    states decreases and the EP2 is shifted to smaller values of $\gamma$. 
    At $J=1$, two $\PT$-broken states emerge directly at $\gamma=0$ with
    degenerate eigenvalues $\mu=1$.
    Thus rendering the ground state unstable for $\gamma\neq0$.
  }%
  \label{fig:linearSpectrum}%
\end{figure}

The transition from the double well to the symmetric triangular case is now
studied in more detail.
The complete analytic expressions for $\gamma\neq0$ do not provide further
insight, therefore we limit our discussion to numerical results starting with
the linear spectrum in Fig.~\ref{fig:linearSpectrum}.
For $J=0$, where the system corresponds to a double-well potential and one
independent well, the ground and excited state of the double well coalesce in
an exceptional point of second order (EP2) where they vanish and two
$\PT$-broken states emerge.
These states have complex conjugate eigenvalues with a finite imaginary part.
The remaining $\PT$-symmetric state is completely localized in the decoupled
well 2, while the double well is empty.
However, any perturbation that adds some number of particles to the wells 1 and
3 still renders the remaining double-well system unstable.

For $J>0$ the structure changes and the ground state stays $\PT$-symmetric for
all values of $\gamma$.
The EP2 however is still present, now occurring between the two excited states
of the three-well system.
For increasing coupling strengths $J$ the EP2 is shifted to lower parameters
$\gamma$ breaking the $\PT$ symmetry of these states and rendering the
system unstable at smaller values of $\gamma$.

For $J=1$ the EP2 reaches $\gamma=0$ where it vanishes due to the Hermiticity
of the system and is replaced by a degeneracy of two eigenvalues. 
For this case, as predicted from perturbation theory, the $\PT$ symmetry
is broken for any value of $\gamma$.

The coupling to an additional well and thus the availability of an
additional channel seems to diminish the capability of the system to support a
stable current.
The net current through the system is given by the particle current from an
external source to the gain well 1.
The continuity equation for this well reads
\begin{equation}
  \frac{\partial}{\partial t}\left|v_1\right|^2 =
    \underbrace{2\gamma\left|v_1\right|^2}_{j_\mrm{ext}}
    -\underbrace{\frac{J}{\rmi}\left(v_1^*v_2-v_1v_2^*\right)}_{j_{12}}
    -\underbrace{\frac{1}{\rmi}\left(v_1^*v_3-v_1v_3^*\right)}_{j_{13}},
  \label{eq:continuity}
\end{equation}
where $v_i$ are the components of the eigenvector and
with the particle currents $j_{ij}$ from well $i$ to $j$ and the
external current of particles into the well $j_\mrm{ext}$.
For stationary states $j_\mrm{ext} = j_{\mrm{12}} + j_{\mrm{13}}$ holds.
At the same time the incoupling in well 1 and the outcoupling in well 3 are
balanced, thus $j_\mrm{ext}$ specifies the net current.

\begin{figure}%
  \centering%
  \includegraphics[width=\columnwidth]{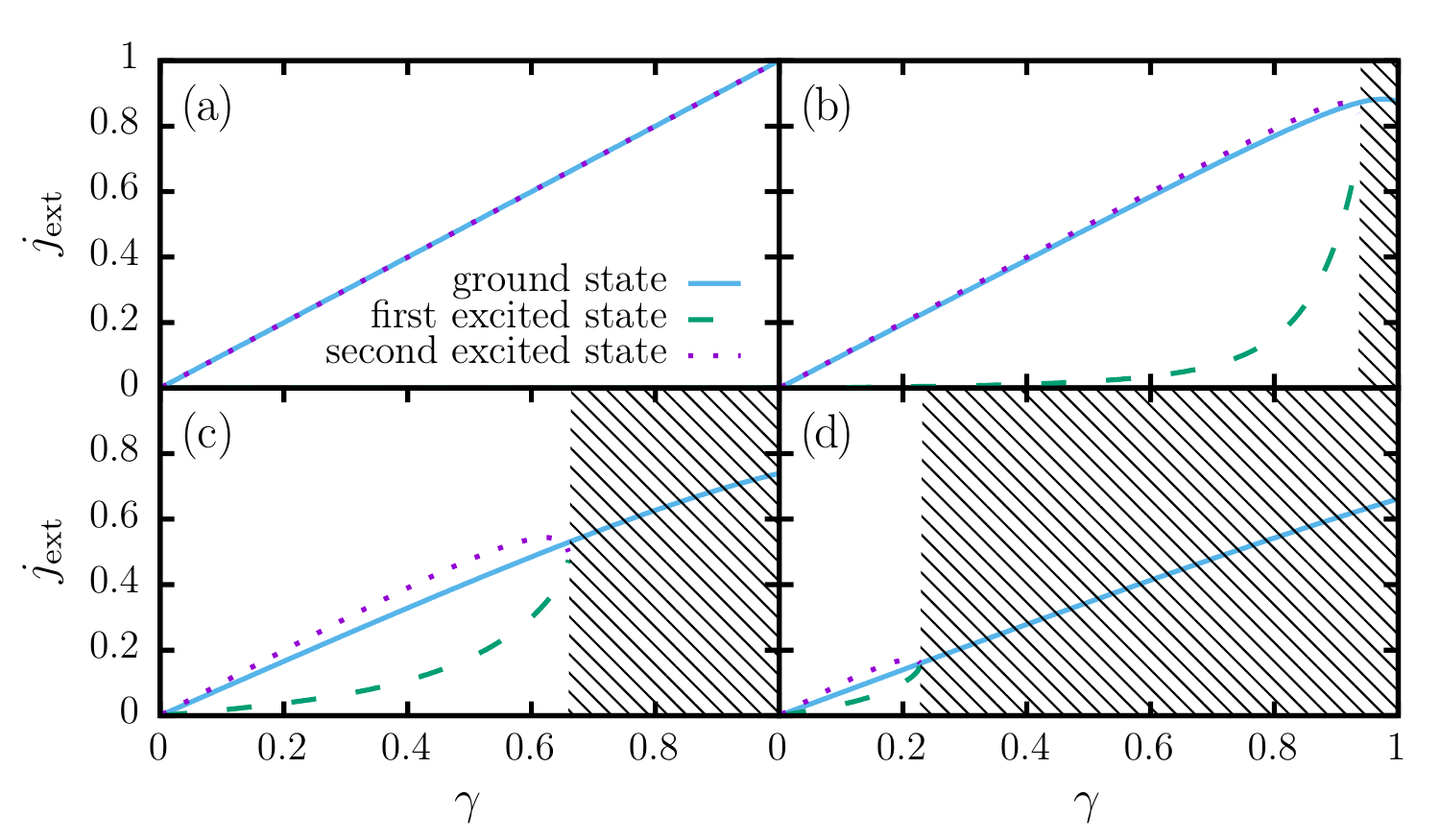}%
  \caption{%
    (Color online)
    Net current through the system $j_\mrm{ext}$ as function of $\gamma$ for
    (a) $J=0$, (b) $J=0.1$, (c) $J=0.4$, (d) $J=0.8$.
    The highlighted areas show regions where $\PT$-broken eigenstates with
    complex eigenvalues exist, thus rendering all stationary states unstable.
    The maximum current, $j_\mrm{ext}=1$, is reachable only for $J=0$.
    For $0<J<1$ the second excited state supports the highest net current
    slightly before the EP.  
  }%
  \label{fig:netCurrent}%
\end{figure}

Figure~\ref{fig:netCurrent} shows the net flow $j_{\mrm{ext}}$ for all three
states and different coupling strengths $J$.
For $J=0$ the ground state and second excited state, which are the solutions
to an equivalent two-mode problem, support equally strong currents.
The maximum current $j_\mrm{ext}=1$ is reached at the exceptional point at
$\gamma=1$.
If the additional well 2 is coupled with $0<J<1$, the ground state is no
longer involved in the EP2 and provides a stationary current even after the two
excited states have vanished.
However in this region the state is no longer stable with respect to any small
perturbations.
The strongest current is achieved by the second excited state for a value of
$\gamma$ slightly smaller than the EP2.
The $\PT$ symmetry breaking at the EP2 is therefore the main reason for the
decrease of the maximal currents as the coupling $J$ is increased.

\begin{figure}%
  \centering%
  \includegraphics[width=\columnwidth]{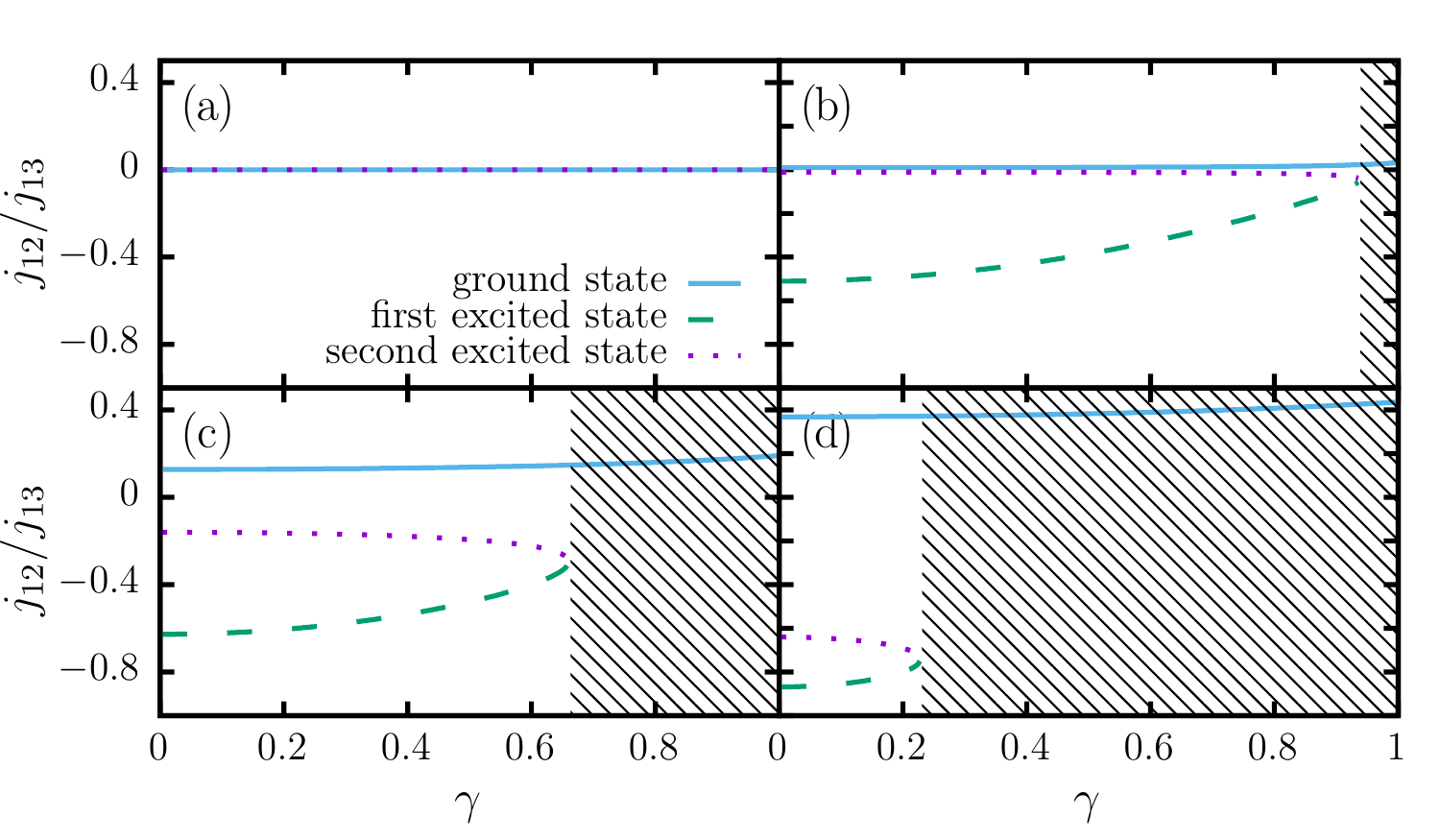}%
  \caption{%
    (Color online)
    Ratio of the current $j_{12}$ through the additional well and
    $j_{13}$ directly to the loss well as function of $\gamma$ for (a) $J=0$,
    (b) $J=0.1$, (c) $J=0.4$, (d) $J=0.8$.
    While the current $j_{13}$ is always positive, negative ratios represent
    negative currents $j_{12}$.
    Again the highlighted areas show regions where $\PT$-broken eigenstates
    exist.
    For $J=0$ no particles can be transferred through well 2.
    After coupling the well to the system the ground state supports a positive
    current through the well, while the excited states show negative currents.
    For $J\to1$ the ratio approaches $-1$.
  }%
  \label{fig:partialCurrents}%
\end{figure}

As a next step we look at the partial particle flows from the gain well 1 to
the loss well 3, $j_{13}$, i.e.\ the direct channel, and to well 2, $j_{12}$,
i.e.\ the additional channel.
These currents are compared in Fig.~\ref{fig:partialCurrents} for different
coupling strengths $J$. 
The ground state is the only stationary state supporting a positive current
through the additional channel.
For increasing coupling parameters $J$ an increasing part of the particles --
for $J=0.8$ (Fig.~\ref{fig:partialCurrents}(d)) -- approximately a third of the
net current, is transported through well 2.

In contrast to this behavior the excited states do not at all support a
positive current through well 2.
Since the current $j_{12}$ is negative, the additional channel transports
particles from the loss to the gain well.
The relative strength of this reverse current increases with higher parameters
$J$, and becomes comparable to the current from the gain to the loss well
$j_{13}$ for $J \to 1$ (Fig.~\ref{fig:partialCurrents}(c),(d)).
For $J \approx 1$ a small parameter $\gamma$ is sufficient to induce a strong
circular current, breaking the $\PT$ symmetry even though only very few
particles enter the system.

Up to now we only discussed linear systems, i.e.\ systems without
inter-particle interaction.
However, in most Bose-Einstein condensates at least a contact interaction,
described by the on-site interaction in Eq.~\eqref{eq:ThreeModeGpe} is present.
In the nonlinear case, the stability of a stationary state $\psi$ is no longer
determined by the complex eigenspectrum but by the linear Bogoliubov-de Gennes
equations:
\begin{subequations}
\label{eq:ThreeModeBdg}
\begin{align}
  \sum\limits_{j=1}^{3} H_{ij} u_j + 2 U \left| \psi_i \right|^2
  u_i + U \psi_i^2 v_i - \mu u_i  &= \omega u_i, \\
  \sum\limits_{j=1}^{3} H^*_{ij} v_j + 2 U \left| \psi_i \right|^2
  v_i + U \psi_i^{*2} u_i - \mu^* v_i &= -\omega v_i,
\end{align}
\end{subequations}
defining the linearized evolution of the perturbation
$\delta\psi_i = u_i \exp{(-\rmi \omega t)} + v_i^* \exp{(\rmi \omega^* t)}$.
If all eigenvalues $\omega$ are real, the state is considered to be stable,
while any non-vanishing imaginary part renders it unstable.

\begin{figure}%
  \centering%
  \includegraphics[width=\columnwidth]{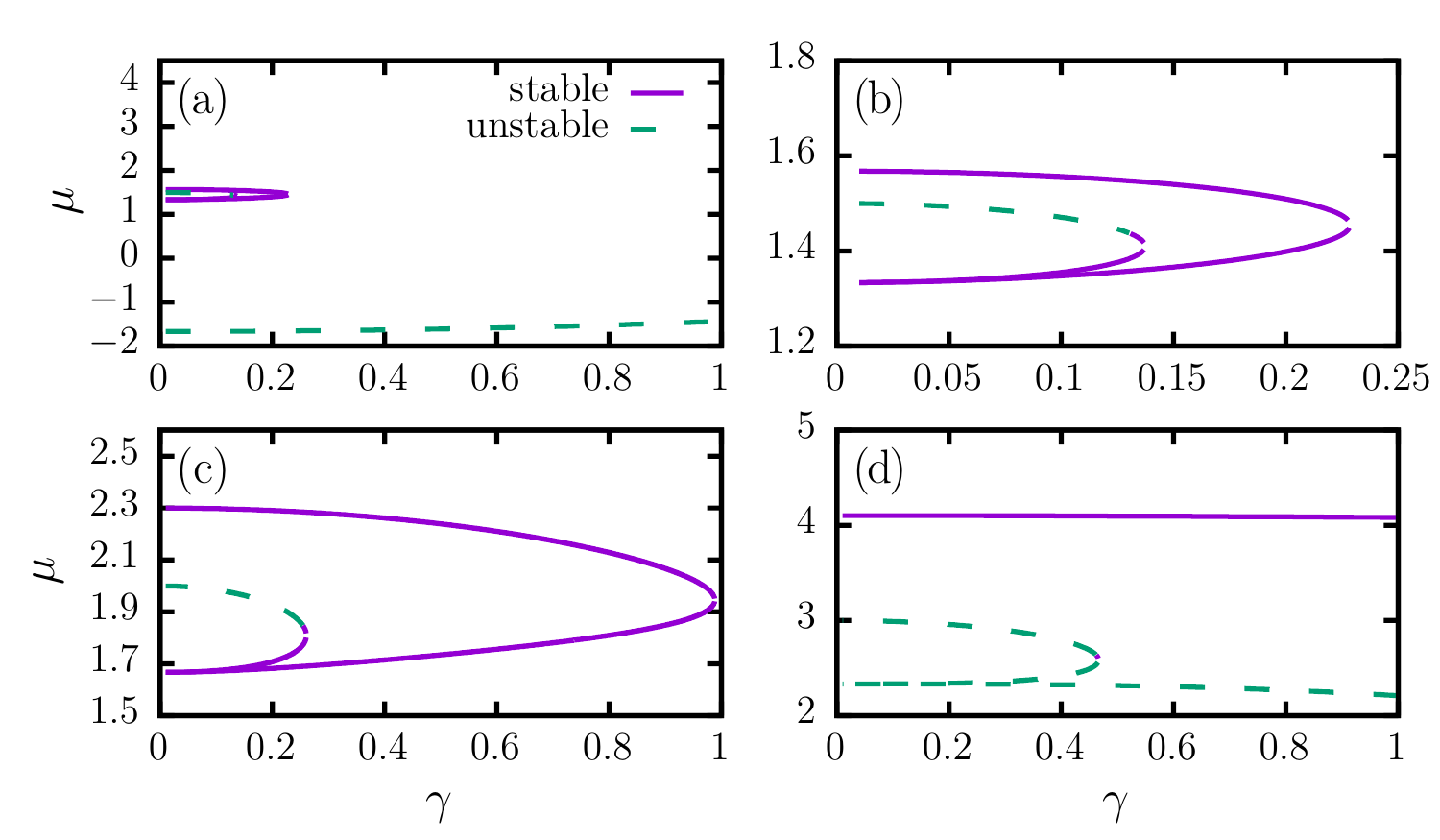}%
  \caption{%
    (Color online)
    Chemical potential $\mu$ of $\PT$-symmetric states against the strength of
    the parameter $\gamma$ for coupling parameter $J=1$ and nonlinearity
    parameter (a),(b) $U=1$, (c) $U=2$, (d) $U=4$.
    Introducing the nonlinearity (a) gives rise to four new states for $\gamma
    \neq 0$, which are shown in the enlarged figures (b),(c),(d). 
    For $U=1$ and $U=2$ three of them are stable (solid lines) and one is
    unstable (dashed line), for $U=4$ all states except the highest excited one
    are unstable.
  }%
  \label{fig:nonlinearSpectrum}%
\end{figure}

Figure~\ref{fig:nonlinearSpectrum} shows the chemical potential for
$J=1$ and different strengths of the nonlinearity $U$.
Again only $\PT$-symmetric states are shown.
The first point to notice is that every nonlinearity strength gives rise to a
set of four new $\PT$-symmetric states
(Fig.~\ref{fig:nonlinearSpectrum}(a),(b)).
Since three of these states are stable, the particle interaction supports a
stationary transport through the system although the symmetry of the system
inhibits all stable currents in the linear case.
The four states vanish in two independent tangent bifurcations.
For a stronger nonlinearity $U=2$ (Fig.~\ref{fig:nonlinearSpectrum}(c)) two
stable states are available up to $\gamma = 1$.
For $U=4$ (Fig.~\ref{fig:nonlinearSpectrum}(d)) only the highest excited
stationary state is still stable.
As already known from in $\PT$-symmetric systems the stability of stationary
states does not change exactly at a bifurcation point but only close to them.
However, in the following discussion we neglect this subtlety.

\begin{figure}%
  \centering%
  \includegraphics[width=\columnwidth]{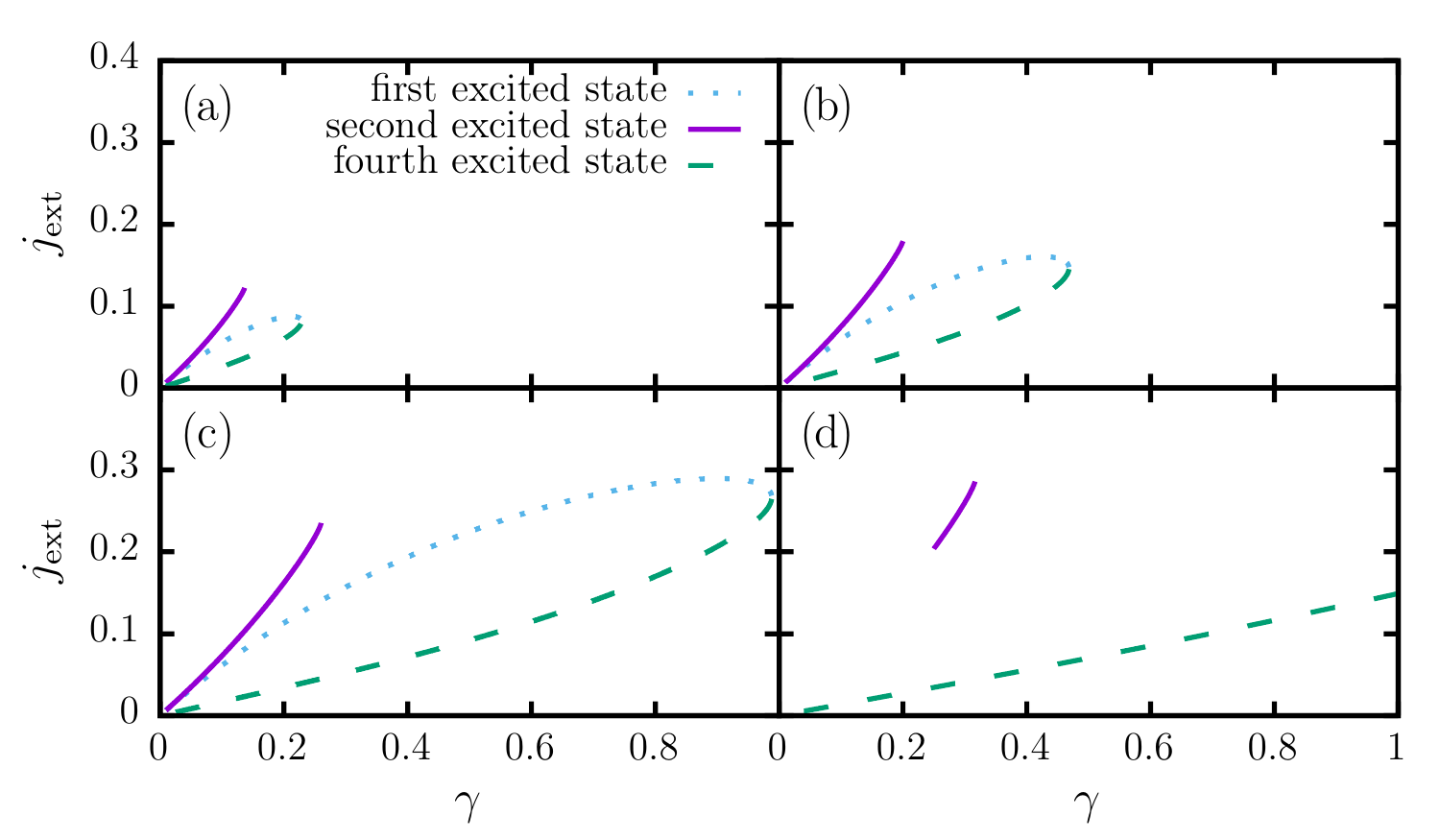}%
  \caption{%
    (Color online)
    Net current through the system given by the incoupling external current
    $j_\mrm{ext}$ against $\gamma$ for (a) $U=1$, (b) $U=1.5$, (c) $U=2$, (d)
    $U=2.5$.
    Only the three stable $\PT$-symmetric states introduced in
    Fig.~\ref{fig:nonlinearSpectrum} are shown.
    For nonlinearities of $U$ up to 2, all three states are able to sustain a
    similar current, which reaches a maximum of approximately $0.3$ at
    $U\approx2$.
    For higher nonlinearities stable states become unstable until only the
    highest excited state remains.
    Its low current is due to the fact that there are only few particles
    present in the gain and loss well. 
  }%
  \label{fig:nonlinearNetCurrent}%
\end{figure}

The three states supporting stationary currents are now discussed in more
detail.
For that purpose, Fig.~\ref{fig:nonlinearNetCurrent} shows the particular net
current for different strengths of the nonlinearity.
For lower nonlinearities the three stable states support a similar maximum net
current but at different values of $\gamma$
(Fig.~\ref{fig:nonlinearNetCurrent}(a)).
This is due to different values of the modulus square of the different wave
functions at wells 1 and 3.
The second excited state is almost fully localized in these wells.
This explains why this state vanishes for lower values of $\gamma$ but is
already supporting a strong current.
For $U\approx2$ the currents reach their maximum of about $0.3$, i.e.\ a third
of the maximum current of the double well with $J=0$.
For higher values of $U$ the first and second excited state become unstable.
The fourth excited state however stays stable since the weak population of the
gain and loss well leads to a weak net current.

\section{Conclusion and outlook}
\label{sec:Conclusion}
Perturbation theory predicts that in linear systems a $\PT$-symmetric
perturbation $\hat{H}_\mrm{P}$ of a Hermitian system leads to real eigenvalues
and therefore stable stationary states if and only if all sets of degenerate
eigenvectors $\{\phi_i\}$ satisfy
$\bra{\phi_i}\hat{H}_\mrm{P}\ket{\phi_j} = 0$.

As a simple example we studied a triple well with one gain, one loss, and one
additional well.
This system violates the above condition in the case of equal coupling
strengths between all wells.
The physical nature of the symmetry breaking was discussed by varying the
coupling strengths while preserving the $\PT$ symmetry.
It was demonstrated that the current supported by the system reaches
its maximum when the third well is not coupled to the system, ergo in the case
of a double well system.
Coupling the well to the system leads to a reverse current, reducing and
ultimately quenching the net current.

While the equally coupled triple well does not support stable currents in the
linear case, introducing a contact interaction gave rise to four new
$\PT$-symmetric states, three of them stable.
These states restore the capability of the system to support a stationary
current.
For an interaction strength above twice the coupling strength between the
wells, the current reaches its maximum of a third of the double-well current in
the linear system.

It would be highly interesting for future studies to analyze spatially
extended three-dimensional triple wells, i.e.\ a full solution in the
three dimensional position space.
While the results from perturbation theory provide good predictions for
linear systems, it is not clear if the inter-particle interaction in
extended potentials is able to restore the support for stationary currents.

\end{document}